\documentclass[12pt]{article}
\pdfoutput=1
\usepackage{jheppub}
\usepackage{bm}
\usepackage{amssymb}
\title{Quantization of area for event and Cauchy horizons of the Kerr--Newman black hole}
\author{Matt Visser}
\affiliation{School of Mathematics, Statistics, and Operations Research, \\
Victoria University of Wellington, PO Box 600, Wellington 6140, New Zealand}
\emailAdd{matt.visser@msor.vuw.ac.nz}
\abstract{
Based on various string theoretic constructions, and various string-inspired generalizations thereof, there have been repeated suggestions that the areas of black hole event horizons might be quantized in a quite specific manner, in terms of linear combinations of square roots of positive integers. It is important to realise that there are significant physical constraints on such integer-based proposals when one (somewhat speculatively) attempts to extend them outside their original extremal and supersymmetric framework.
Specifically, in their most natural and direct physical interpretations,  some of the more speculative integer-based proposals for the quantization of horizon areas \emph{fail} for the ordinary Kerr--Newman black holes in (3+1) dimensions, \emph{essentially because the fine structure constant is not an integer}. 
A more baroque interpretation involves asserting the fine structure constant is the square root of a rational number; but such a proposal has its own problems.
Insofar as one takes (3+1) general relativity (plus the usual quantization of angular momentum and electric charge) as being paramount, the known explicitly calculable spectra of horizon areas for the physically compelling Kerr--Newman spacetimes indicate that some caution is called for when assessing the universality of some of the more speculative integer-based string-inspired proposals.

\bigskip
\noindent
arXiv:1204.3138 [gr-qc] --- 
14 April 2012; 21 April 2012; 8 May 2012; \\
\LaTeX-ed \today
}
\begin{document}
%---------------------------------------------------------------------------------------------------------------------------------------------
\maketitle
%------------------------------------------------------------------------------------------------------------------------------------------
% Basic definitions
%------------------------------------------------------------------------------------------------------------------------------------------
%------------------------------------------------------------------------------------------------------------------------------------------
%------------------------------------------------------------------------------------------------------------------------------------------
\def\R{{\mathbb{R}}}
\def\N{{\mathbb{N}}}
\def\Z{{\mathbb{Z}}}
\def\Q{{\mathbb{Q}}}

%------------------------------------------------------------------------------------------------------------------------------------------
%---------------------------------------------------------------------------------------------------------------------------------------------
\section{Introduction}
%---------------------------------------------------------------------------------------------------------------------------------------------
\label{S:intro}
%---------------------------------------------------------------------------------------------------------------------------------------------

Various string-theoretic and string-inspired constructions have lead to the suggestion that black hole event horizon areas might follow the quantization rule~\cite{Horowitz:1996, KeskiVakkuri:1996, Horowitz:1996b, Halyo:1996, Horowitz:1996c, Larsen:1997, Cvetic:1997, Cvetic:2009, Cvetic:2010, Galli:2011}
\begin{equation}
A_+ = 8\pi L_P^2 \left\{  \sqrt{N_1} + \sqrt{N_2}  \,\right\}; \qquad N_1,N_2 \in \mathbb{N}.
\label{E:1}
\end{equation}
Such a quantization rule was first developed in the context of extremal and supersymmetric black holes where the quantities $N$ are known to quite literally be integers. Even for near-extremal black holes one should be more careful, with reference~\cite{Horowitz:1996b} for instance characterizing the $N$ in the following manner:  ``we will refer to them as the numbers of branes, antibranes and strings because (as will be seen) they reduce to those numbers in certain limits where these concepts are well defined''. The relevant calculations are often best described as string-inspired rather than string-derived.  For instance reference~\cite{Larsen:1997} describes the calculational framework as ``an effective low energy description of black holes'', not a general string theory construction \emph{per se}. 
These qualifications and reservations have sometimes been neglected in the subsequent literature, and can then lead to a somewhat misleading view of the situation.

In situations where there is both an inner (Cauchy) horizon and outer (event) horizon one sometimes encounters the stronger string-inspired conjecture that~\cite{Larsen:1997, Cvetic:1997, Cvetic:2009,  Cvetic:2010, Galli:2011, Meessen:2012, Castro:2012}
\begin{equation}
A_+ A_-  = (8\pi L_P^2)^2 N; \qquad N \in \mathbb{N}.
\label{E:2}
\end{equation}
A more cautious interpretation is that the $N$ might be products of generalized charges, rather than integers~\cite{Cvetic:2010}. For instance in reference~\cite{Castro:2012} one finds the cautionary comment: ``The precise statement is that the products of areas is independent of the mass of the black hole and therefore depends solely on the quantized charges.'' Such careful cautionary comments are sometimes omitted in other parts of the literature, and the $N$ are sometimes naively taken to be integral without further comment. 
If one were to take the $N$ to \emph{literally} be integers than this would imply~\cite{Larsen:1997, Cvetic:1997, Cvetic:2009,   Cvetic:2010, Galli:2011, Meessen:2012, Castro:2012}
\begin{equation}
A_\pm = 8\pi L_P^2 \left\{  \sqrt{N_1} \pm \sqrt{N_2} \,  \right\}; \qquad N_1,N_2 \in \mathbb{N}.
\label{E:3}
\end{equation}
These integer-based string-inspired proposals are then sometimes conjectured to have a \emph{universal} and \emph{literal} validity, for integer $N$, far beyond the realm in which they were originally obtained. It is this conjecture of universal and literal validity of the integer-based proposals which will be addressed in this current article --- and we shall see that there are good reasons for being somewhat cautious in this regard.

Specifically, when attempting to connect these specific string-inspired proposals to black holes in (3+1) dimensions, such as the usual Kerr--Newman (electrically charged and rotating) black hole~\cite{MTW, Wald, Newman:1965a, Newman:1965b}, one should take particular care to note that while angular momentum is naturally quantized in terms of $\hbar$, electric charges are instead quantized in terms of the electron charge $Q_e$ --- and it is through the combination $\alpha = Q_e^2/\hbar c$ that the physical fine structure constant $\alpha\approx 1/137.035999074(44)$ will be seen to enter the formulae for the areas of the inner and outer horizons of Kerr--Newman black holes.

If one is to take the proposed formulae (\ref{E:1})--(\ref{E:3}) literally, we shall soon see that (if the $N$ are literally to be integers) then for the fine structure constant we would need to set $\alpha=2$, a grossly unphysical value. If we relax the string-inspired proposals by ``merely'' assuming that the product of horizon areas in equation (\ref{E:2}) be some rational multiple of $(8\pi L_P^2)^2$, then we would still at the very least need the physical fine structure constant to be the square root of some rational number --- $\alpha^2\in \Q$. Any such constraint, apart from being rather baroque and certainly unexpected,  is certainly difficult to reconcile with the known renormalization group flow of the fine structure constant. Additionally, such speculative demands on the rational nature of $\alpha^2$ rapidly begin to take on the flavour of Eddington's ``fundamental theory''~\cite{Eddington}. 

A physically more reasonable  interpretation is simply to realise that the areas of Kerr--Newman horizons, while certainly quantized, are not quantized in quite as simple a manner as some of the more speculative integer-based string-inspired constructions suggest. For Kerr--Newman black holes we shall derive a \emph{non-integral} quantization formula for the product of horizon areas in terms of spin, electric charge, and the fine structure constant.  Related intrinsically \emph{non-integral} quantization formulae for the individual inner (Cauchy) and outer (event) horizon will also be obtained.

That is: Insofar as currently available string-inspired conjectures  lead to integer-quantized products of horizon areas, these specific conjectures do not seem to be in any straightforward manner compatible with ordinary (3+1) general relativity. Conversely, insofar as one takes ordinary (3+1) general relativity (plus quantization of angular momentum and electric charge) as being paramount, the spectra of horizon areas does not resemble that of some of the more speculative integer-based string-inspired theoretical conjectures. It seems that  some caution is called for when assessing the universality of some of the more speculative integer-based string-inspired proposals.

%---------------------------------------------------------------------------------------------------------------------------------------------
\section{Kerr--Newman black holes}
%---------------------------------------------------------------------------------------------------------------------------------------------
\label{S:KN}
%---------------------------------------------------------------------------------------------------------------------------------------------

In Boyer--Lindquist coordinates the (3+1) Kerr--Newman geometry of a rotating charged black hole takes the form~\cite{MTW, Wald, Newman:1965a, Newman:1965b} 
\begin{equation}
ds^2 = - {\Delta\over\rho^2}(dt - a \sin^2\theta \, d\phi)^2 + \rho^2\left({dr^2\over\Delta}+d\theta^2\right)
+{\sin^2\theta\over\rho^2} \left((r^2+a^2)d\phi - a \, dt\right)^2,
\end{equation}
where
\begin{equation}
a=J/m; \qquad \rho^2 = r^2 + a^2 \cos^2\theta; \qquad \Delta = r^2 - 2m r + a^2 + Q^2;
\end{equation}
and we are using standard geometrodynamic units ($G=c=1$, as in MTW~\cite{MTW}). 
For a discussion of the uncharged rotating Kerr black hole see~\cite{Kerr:1963, Kerr:book, Kerr:survey}. 
The horizons, where $\Delta=0$, are located at
\begin{equation}
r_\pm =  m \pm \sqrt{m^2 - a^2 - Q^2}.
\end{equation}
The areas of these horizons are easily seen to be
\begin{equation}
A_\pm = 4\pi  (r_\pm^2 + a^2 ) = 4\pi \left\{ 2m^2 - Q^2 \pm 2 m   \sqrt{m^2 - a^2 - Q^2} \right\}.
\end{equation}
Then
\begin{eqnarray}
A_+ A_- &=& (4\pi)^2 [ (2m^2 - Q^2)^2 - (2m)^2(m^2 - a^2 - Q^2) ]
\\
&=& (4\pi)^2 [ (4 m^2 - 4 m^2 Q^2 + Q^4) - (4m^4 - 4 m^2 a^2 - 4 m^2 Q^2) ]
\\
&=& (4\pi)^2 [ 4 m^2 a^2  + Q^4 ].
\end{eqnarray}
That is, (in geometrodynamic units) 
\begin{equation}
A_+A_-  = (8\pi)^2 \left[ J^2  + {Q^4\over4} \right].
\label{E:product}
\end{equation}
The remarkable feature of this purely classical result is that this product is independent of the mass $m$ of the black hole, and depends only on the conserved charges $J$ and $Q$. 
For extensive discussion of this purely classical (3+1) dimensional result and its extensions see for instance~\cite{Ansorg:2010, Ansorg:2009, Ansorg:2008, Hennig:2009}. Similar purely classical results for the product of inner and outer horizon areas are known to hold for a wide variety of more exotic black holes in four or more dimensions~\cite{Larsen:1997, Cvetic:1997, Cvetic:2009, Cvetic:2010, Galli:2011, Meessen:2012, Castro:2012}, with the product depending on various combinations of the relevant angular momenta and charges (moduli), and quite often being independent of the black hole mass.

%---------------------------------------------------------------------------------------------------------------------------------------------
\section{Quantized charge and angular momentum}
%---------------------------------------------------------------------------------------------------------------------------------------------
\label{S:quantized}
%---------------------------------------------------------------------------------------------------------------------------------------------

In geometrodynamic units the electron charge is $Q_e = \sqrt{\alpha} \, M_P =  \sqrt{\alpha} \, L_P$. (Similarly, in Planck units~\cite{Planck:1899} where $G=\hbar=c=1$ the electron charge is $Q_e = \sqrt{\alpha}$. 
In contrast, in so called Stoney units~\cite{Stoney:1881, Stoney:1883}, which pre-date Planck units by some 20 years, one has $G=c=Q_e=1$, but then $\hbar = 1/\alpha$. Either way there is an $\alpha$-dependent mismatch between the Planck constant and the charge on the electron.)
In physical units the formula (\ref{E:product}) becomes
\begin{equation}
A_+A_-  = (8\pi L_P^2)^2 \left[ {J^2\over\hbar^2}   + {\alpha^2 Q^4\over4 Q_e^4} \right].
\end{equation}
If we take the angular momentum to be quantized in units of $\hbar$, (or ${1\over2}\hbar$ for fermionic black holes), and the electric charge to be quantized in terms of the electron charge $Q_e$,  then
\begin{equation}
J^2 \to  j(j+1) \hbar^2; \quad j \in \N/2; \qquad  Q \to q Q_e; \quad q \in \Z;
\end{equation}
and so
\begin{equation}
A_+A_-  = (8\pi L_P^2)^2 \left[  j(j+1)   + {\alpha^2 q^4\over4} \right]; \qquad j \in \N/2; \quad q \in \Z.
\end{equation}
This simple calculation is enough to verify our key claims --- for Kerr--Newman black holes the product of areas is \emph{not} quantized as simple integer multiples of $(8\pi L_P)^2$. While under the stated assumptions the product of horizon areas is certainly quantized, the quantization formula contains what is from the point of view of general relativity a free parameter --- the fine structure constant. 

As presaged in the introduction, if we demand strict and literal compatibility with equation (\ref{E:2}), then for integer spin this product of areas will not be an integer multiple of $(8\pi L_P^2)^2 $ \emph{unless} the fine structure constant is set to the grossly unphysical value of $\alpha =2$, to be compared with the physical value of $\alpha\approx1/137.035999074(44)$. For half-integer spin we can never obtain strict and literal compatibility with equation (\ref{E:2}), though we could instead set $\alpha=1$ and so obtain integer multiples of the reduced quantity $(4\pi L_P^2)^2$.

Furthermore, we note that this product of areas will not even be a \emph{rational} multiple of $(8\pi L_P^2)^2 $ unless $\alpha$ is the square root of a rational number --- $\alpha^2 \in \Q$ --- this is a rather baroque constraint which is difficult to square with the known renormalization group running of $\alpha$, which implies that as a function of energy $\alpha$ will generically flow through both rational and irrational and even transcendental numbers. Indeed, speculative demands on the rational nature of $\alpha^2$ rapidly begin to take on the flavour of Eddington's ``fundamental theory''~\cite{Eddington}.

%---------------------------------------------------------------------------------------------------------------------------------------------
\section{Individual horizon areas}
%---------------------------------------------------------------------------------------------------------------------------------------------
\label{S:individual}
%---------------------------------------------------------------------------------------------------------------------------------------------

For the inner and outer horizons of the Kerr--Newman geometry we have (in physical units) the mass-independent bounds
\begin{equation}
A_+  \geq 8\pi L_P^2 \; \sqrt{  j(j+1)   + {\alpha^2 q^4\over 4} } \geq A_-. 
\label{E:bound}
\end{equation}
If in addition we permit one to explicitly use information concerning the mass, then for the inner and outer horizons of the Kerr--Newman geometry we have
 (in geometrodynamic units) the individual equalities
\begin{equation}
A_\pm = 8\pi \left\{ m^2 - {Q^2\over2}  \pm   \sqrt{m^4 - m^2 Q^2 - J^2} \right\},
\end{equation}
or equivalently
\begin{equation}
A_\pm = 8\pi \left\{ m^2 - {Q^2\over2}  \pm   \sqrt{\left(m^2 -{Q^2\over2}\right)^2 - {Q^4\over 4}  - J^2} \right\}.
\end{equation}
In physical units this becomes 
\begin{eqnarray}
A_\pm &=& 8\pi L_P^2 \Bigg\{ {m^2\over m_P^2} - {\alpha \; q^2\over2} 
 \pm  \sqrt{{m^4\over m_P^4}  -  {m^2\over m_P^2}{\alpha \; q^2} - j(j+1) } \Bigg\},
\end{eqnarray}
or equivalently
\begin{eqnarray}
A_\pm &=& 8\pi L_P^2 \Bigg\{ \left[{m^2\over m_P^2} - {\alpha \; q^2\over2}\right]
 \pm  \sqrt{\left[{m^2\over m_P^2}  - {\alpha \; q^2\over2}\right]^2  -  {\alpha^2 \; q^4\over 4} - j(j+1) } \bigg\}.
\end{eqnarray}
In general this is the best one can do;  \emph{without other external assumptions we have no particular reason for both $m/m_P$ and the fine structure constant $\alpha$ not being generic real numbers}.  

We have already seen that to (exactly and literally) match this spectrum with equation (\ref{E:2}) we would need to demand the extremely unphysical condition $\alpha=2$. To additionally literally match this with equations (\ref{E:1}) and   (\ref{E:3}) we would then need to additionally demand 
\begin{equation}
m^2 = m_P^2 \left\{ \sqrt{N_1} + q^2 \right\};  \qquad N_1 = N_2 +j(j+1)+q^4.
\end{equation}
While this last step is not particularly problematic, it is the first condition $\alpha=2$ that shows this is not a physically viable proposition.

Furthermore, even if one makes the very strong (and as we have indicated above, physically rather dubious) assumption that \emph{both} of the quantities $m/m_P$ and $\alpha$ are rational numbers, (not even just the square roots of rational numbers), one would still not obtain the integer-based string-inspired ansatze of equations (\ref{E:1})--(\ref{E:3}). Even under these very strong assumptions the best one could possibly hope for is the following \emph{rational generalization} of the usual string inspired ansatze:
\begin{equation}
A_\pm = 8\pi L_P^2 \left\{  \sqrt{N_1\over M_1} \pm \sqrt{N_2\over M_2} \,  \right\}; \qquad N_1,N_2, M_1, M_2 \in \mathbb{N};
\end{equation}
and
\begin{equation}
A_+ A_- = (8\pi L_P^2)^2 \left\{  {N_1\over M_1}-  {N_2\over M_2} \,  \right\}; \qquad N_1,N_2, M_1, M_2 \in \mathbb{N}.
\end{equation}
This can be rewritten as
\begin{equation}
A_\pm = {8\pi L_P^2\over M_1 M_2} \left\{  \sqrt{N_1M_1 M_2^2} \pm \sqrt{N_2 M_2 M_1^2} \,  \right\}; \qquad N_1,N_2, M_1, M_2 \in \mathbb{N};
\end{equation}
and
\begin{equation}
A_+ A_- = \left({8\pi L_P^2\over M_1 M_2}\right)^2 \left\{  {N_1M_1 M_2^2}-  {N_2 M_2 M_1^2} \,  \right\}; \qquad N_1,N_2, M_1, M_2 \in \mathbb{N}.
\end{equation}
Overall, this has the net effect of somewhat resembling the string-inspired quantization rules of equations  (\ref{E:1})--(\ref{E:3}), but now in terms of the ``reduced'' Planck area $L_P^2\to L_P^2/(M_1 M_2)$.  Even then, one is stepping well outside the original proposal of equations  (\ref{E:1})--(\ref{E:3}), and one should be very aware of the strong assumptions one is making, ($m/m_P \in \Q$, $\alpha \in \Q$), assumptions that have no real independent justification, apart from the desire to somehow \emph{force} the Kerr--Newman black hole into a vaugely string-inspired framework.

%---------------------------------------------------------------------------------------------------------------------------------------------
\section{Avoiding naked singularities}
%---------------------------------------------------------------------------------------------------------------------------------------------
\label{S:naked}
%---------------------------------------------------------------------------------------------------------------------------------------------

As an internal consistency check note that the condition for having horizons (either extremal or non-extremal) and avoiding naked singularities is most often phrased as 
\begin{equation}
m^2 \geq a^2 + Q^2. 
\end{equation}
When working in terms of $J$ rather than $a$ it is better to rephrase this as
\begin{equation}
m^4 - m^2 Q^2 - J^2 \geq 0.
\end{equation}
The physically relevant  solution to this inequality is (in geometrodynamic units)
\begin{equation}
m^2 \geq   {Q^2 + \sqrt{Q^4+4J^2}\over 2},
\end{equation}
whence (in physical units) the condition for the existence of a horizon becomes a bound on the mass:
\begin{equation}
m^2 \geq m_P^2 \left\{  \sqrt{j(j+1) + {\alpha^2 \; q^2\over4} } 
+ {\alpha \; q^2\over2}\right\}.
\end{equation}
In this situation the area of the outer horizon is bounded by 
\begin{equation}
A_+ \geq  4\pi \{ 2m^2 - Q^2\} \geq 8\pi \sqrt{J^2+{Q^2\over 4}},
\end{equation}
or in physical units
\begin{equation}
A_+  \geq 8\pi L_p^2 \; \sqrt{j(j+1) + {\alpha^2 \; q^4\over4} },
\end{equation}
consistent with the bound  we had previously given in equation (\ref{E:bound}).

%---------------------------------------------------------------------------------------------------------------------------------------------
\section{Bekenstein-style area quantization}
%---------------------------------------------------------------------------------------------------------------------------------------------
\label{S:Bekenstein}
%---------------------------------------------------------------------------------------------------------------------------------------------

Predating the string-inspired conjectures of equations (\ref{E:1})--(\ref{E:3}) by some 25 years, the original Bekenstein proposals for black hole entropy~\cite{Bekenstein:1973, Bekenstein:1974} amount to arguing that the event horizon (outer horizon) should instead satisfy 
\begin{equation}
A_+ = 8\pi L_P^2 \; \zeta\; N ; \qquad N \in \mathbb{N},
\label{E:B1}
\end{equation}
where $\zeta$ is a dimensionless number of order unity. 
Note that Bekenstein's proposal is qualitatively very different from the string-inspired ansatze of equations (\ref{E:1})--(\ref{E:3}). If we insert Bekenstein's proposal for the quantization of event horizon areas into the explicit formula for $A_+$ we obtain a rather specific mass spectrum:
\begin{equation}
m^2 = m_P^2 \left\{ {\zeta N\over2} +   {\alpha\; q^2\over2} 
 + {j(j+1) + {1\over4} \alpha^2 \; q^4 \over 2\zeta N}  \right\}.
\end{equation}
Here $N$ and $j$ are both natural numbers (non-negative integers),  $q$ is an integer, while generically we can say no more than $\alpha\in\R$ and $\zeta=O(1)$.  This spectrum is nowhere near as straightforward as being simply integer spaced, or even square-root-of-integer spaced, and the special case of the Schwarzschild black hole (with $m = m_P \sqrt{\zeta N/2}\,$)  is rather misleading in this regard. Furthermore, if one considers the inner horizon then under the Bekenstein quantization ansatz one has
\begin{equation}
A_-  = {8\pi L_P^2\over \zeta N} \left[  j(j+1)   +{1\over4} \alpha^2 \; q^4\right]. 
\end{equation}
While this spectrum is certainly discrete, it is nowhere near as simple as being integer spaced.  Generically, it is not even rational spaced.

%---------------------------------------------------------------------------------------------------------------------------------------------
\section{Discussion}
%---------------------------------------------------------------------------------------------------------------------------------------------
\label{S:disc}
%---------------------------------------------------------------------------------------------------------------------------------------------

We have seen that for the physically interesting Kerr--Newman black holes in (3+1) dimensions it is possible to derive a \emph{non-integral} horizon-area quantization formula (for the product of inner and outer horizon areas) in terms of spin, electric charge, and the fine structure constant, a formula that is independent of the underlying black hole mass and depends only on the conserved charges. For inner and outer horizons considered separately, there is still a  \emph{non-integral} horizon-area quantization formula, now in terms of  spin, electric charge,  the fine structure constant, and the mass. The appearance of the fine structure constant in these formulae is perhaps unexpected, though it should not be, and difficult to reconcile with the more naive of the string-inspired conjectures.  Explicitly keeping track of occurrences of the fine structure constant in the relevant formulae is more than just a cosmetic detail --- it carries real physical impact and forces one to think carefully about the underlying physics. 

In the appendices we briefly sketch the complications attendant on introducing magnetic charges, fractional (quark) charges, and multiple $U(1)$ gauge fields. The situation does not qualitatively improve. 
Specifically: Insofar as currently available string-inspired conjectures lead to integer-quantized products of  horizon areas, then these specific integer-based conjectures do not seem compatible with any straightforward application of (3+1) general relativity; insofar as one takes (3+1) general relativity (Kerr--Newman spacetime plus quantization of angular momentum and charge) as being paramount, the resulting spectra of horizon areas does not resemble those of the more naive integer-based string-inspired conjectures. Considerable caution is called for when assessing the universality of some of the more speculative integer-based string-inspired proposals.

%---------------------------------------------------------------------------------------------------------------------------------------------
\section*{Acknowledgments} 
%---------------------------------------------------------------------------------------------------------------------------------------------

This research was supported by the Marsden Fund, 
administered by the Royal Society of New Zealand.  

%---------------------------------------------------------------------------------------------------------------------------------------------
\appendix
%---------------------------------------------------------------------------------------------------------------------------------------------
\section{The effect of including magnetic charge}
%---------------------------------------------------------------------------------------------------------------------------------------------
\label{S:magnetic}
%---------------------------------------------------------------------------------------------------------------------------------------------

It is possible to generalize the discussion of this article to include magnetic as well as electric charges, but there is no great gain in doing so. The formulae for horizon area are somewhat more complicated, but the qualitative features of the preceding discussion remain unaltered. In this appendix we sketch several key features of this generalization.

The starting point is to realise that the Kerr--Newman spacetime (like the non-rotating Reissner--Nordstrom spacetime) does not really distinguish the effects of electric charge from magnetic charge,  and in the formula for the spacetime metric one should really write
\begin{equation}
Q^2 = Q^2_{electric} + Q^2_{magnetic}.
\end{equation}
The standard Dirac quantization condition (in particle physics units) reads~\cite{Dirac}
\begin{equation}
{Q_{electric} \; Q_{magnetic} \over \hbar c} = {g\over 2};  \qquad g \in \Z.
\end{equation}
This implies the quantization of magnetic charge. 
If we take $Q_e$ to be the electron charge and $Q_m$ to be the smallest possible non-zero magnetic charge, then we have  $Q_{electric} = q Q_e$ while $Q_{magnetic} = q_m \, Q_m$, and
\begin{equation}
{Q_e \; Q_m\over \hbar c} = {1\over 2}.
\end{equation}
This is perhaps more usefully written in terms of the ``magnetic fine structure constant'' $\alpha_m = Q_m^2/\hbar c$ as
\begin{equation}
\alpha \; \alpha_{m} = {1\over 4}.
\end{equation}
Returning to our area formulae we now substitute (in geometrodynamic units)
\begin{equation}
Q^2 =  q^2 Q_e^2 + q_m^2 Q_m^2;  \qquad q, q_m \in \Z, 
\end{equation}
whence in physical units
\begin{equation}
Q^2 \to  \alpha q^2 + \alpha_m q_m^2 = \alpha q^2 + {q_m^2\over 4\alpha} = {1\over2} \left\{ 2\alpha q^2 + {q_m^2\over 2\alpha} \right\} ;  \qquad q, q_m \in \Z.
\end{equation}
The net result is that to include the possibility of magnetic charge we need merely substitute
\begin{equation}
\alpha q^2 \to  \alpha q^2 + \alpha_m q_m^2 = \alpha q^2 + {q_m^2\over 4\alpha} = {1\over2} \left\{ 2\alpha q^2 + {q_m^2\over 2\alpha} \right\} 
\end{equation}
in the various formulae appearing in the main text of the article.
In particular
\begin{equation}
A_+A_-  = (8\pi L_P^2)^2 \left[  j(j+1)   + {(\alpha q^2+\alpha_m q_m^2)^2  \over4} \right],
\end{equation}
which is equivalent to the more symmetrical looking form
\begin{equation}
A_+A_-  = (8\pi L_P^2)^2 \left[  j(j+1)   + {1\over16} \left(2\alpha q^2+{q_m^2\over2\alpha} \right)^2\right]. 
\end{equation}
This formula will add complications to the previous discussion, but will not improve the (non-)matching between the explicitly known Kerr--Newman horizons and the integer-based string-inspired ansatz of equation (\ref{E:2}). 

For the inner and outer horizons of the Kerr--Newman geometry we now have (in physical units) the mass-independent bounds
\begin{equation}
A_+  \geq 8\pi L_P^2 \; \sqrt{  j(j+1)   + {1\over16} \left(2\alpha q^2+{q_m^2\over2\alpha} \right)^2} \geq A_-. 
\label{E:bound:A}
\end{equation}
If we additionally permit one to use information concerning the mass, then 
\begin{eqnarray}
A_\pm &=& 8\pi L_P^2 \Bigg\{ {m^2\over m_P^2} - {1\over 4}\left(2\alpha q^2+{q_m^2\over2\alpha} \right)
%\\
%&&
%\qquad\quad
 \pm  \sqrt{{m^4\over m_P^4}  -  {1\over2}{m^2\over m_P^2}\left(2\alpha q^2+{q_m^2\over2\alpha} \right) - j(j+1) } \Bigg\},
 \nonumber
 \\
\end{eqnarray}
or equivalently
\begin{eqnarray}
A_\pm &=& 8\pi L_P^2 \Bigg\{  \left[{m^2\over m_P^2} - {1\over4}\left(2\alpha q^2+{q_m^2\over2\alpha} \right)\right]
\\
&&
\qquad\quad
 \pm  \sqrt{\left[{m^2\over m_P^2}  - {1\over4}\left(2\alpha q^2+{q_m^2\over2\alpha} \right)\right]^2  -  {1\over16}\left(2\alpha q^2+{q_m^2\over2\alpha}\right)^2 - j(j+1) } \bigg\}.
 \nonumber
\end{eqnarray}
Again, this formula will add complications to the previous discussion, but will not improve the (non-)matching between the explicitly known Kerr--Newman horizons and the integer-based string-inspired ansatz of equation (\ref{E:3}).

The condition for the existence of a horizon (the absence of a naked singularity) now becomes  (in physical units) the bound on the mass:
\begin{equation}
m^2 \geq m_P^2 \left\{  \sqrt{j(j+1) + {1\over16} \left(2\alpha q^2+{q_m^2\over2\alpha} \right)^2} 
+ {1\over4}\left(2\alpha q^2+{q_m^2\over2\alpha} \right) \right\}.
\end{equation}

If we now invoke Bekenstein's proposal for the quantization of event horizon areas, and insert it  into the explicit formula for $A_+$  in the presence of magnetic charge, we obtain the rather specific mass spectrum
\begin{equation}
m^2 = m_P^2 \left\{ {\zeta N\over2} +   {1\over4} \left(2\alpha q^2+{q_m^2\over2\alpha} \right)   
 + {1  \over 2\zeta N} \left[j(j+1) + {1\over16} \left(2\alpha q^2+{q_m^2\over2\alpha} \right)^2\right] \right\},
\end{equation}
where again  $N$ and $j$ are both natural numbers, and $q$ and $q_m$ are integers, while generically we can say no more than $\alpha\in\R$ and $\zeta=O(1)$. Furthermore, if one considers the inner horizon then under the Bekenstein quantization ansatz one has
\begin{equation}
A_-  = {8\pi L_P^2\over \zeta N} \left[  j(j+1)   +{1\over16} \left(2\alpha q^2+{q_m^2\over2\alpha} \right)^2\right]. 
\end{equation}
Again, while this would add complications to the previous discussion, it will not improve the (non-)matching between the explicitly known Kerr--Newman results and the integer-based string-inspired ansatze of equations (\ref{E:1})--(\ref{E:3}). 

%---------------------------------------------------------------------------------------------------------------------------------------------
\section{Fractional (quark) charge}
%---------------------------------------------------------------------------------------------------------------------------------------------
\label{S:quark}
%---------------------------------------------------------------------------------------------------------------------------------------------

The obvious change when one deals with fractional (quark) charge is that now $q\in\Z/3$. Less obvious is what one should do with the Dirac quantization condition. There is widespread (but perhaps not universal) agreement to the effect that only integer (non-confined) electric charges are subject to Dirac quantization so that $Q_m$ and $\alpha_m$ are not affected. One need then also decide whether magnetic monopole charges can come in (confined) fractions, $q_m\in \Z/3$? Fortunately none of these issues qualitatively affect the details of the preceding discussion.  As long as both $q$ and $q_m$ are discrete integer multiples of \emph{some} ``smallest non-zero quantity'' the arguments presented above will automatically follow.

%---------------------------------------------------------------------------------------------------------------------------------------------
\section{Multiple $U(1)$ gauge fields}
%---------------------------------------------------------------------------------------------------------------------------------------------
\label{S:multiple}
%---------------------------------------------------------------------------------------------------------------------------------------------

In the same way that the Kerr--Newman spacetime is insensitive to whether the source of the electromagnetic field is an electric or a magnetic charge one could add arbitrary multiple $U(1)$ gauge fields with Maxwell-like Lagrangians. All that would happen is that in the spacetime metric one would need to replace
\begin{equation}
Q^2 \to \sum_i   \left\{ Q_{electric,i}^2 + Q_{magnetic,i}^2 \right\}.
\end{equation}
In physical units this would become
\begin{equation}
Q^2 \to \sum_i  \left\{ \alpha_i \, q_i^2 + \alpha_{m,i} \, q_{m,i}^2 \right\},
\end{equation}
where one is now summing over multiple ``electromagnetic'' fields with multiple electric and magnetic fine structure constants. One might now speculate that with many $U(1)$ fields and many carefully chosen charges one might somehow arrange (or rather force) the horizon areas to be integer-quantized in the sense of equation (\ref{E:2}) by demanding 
\begin{equation}
{1\over2} \; \sum_i  \left\{ \alpha_i \, q_i^2 + \alpha_{m,i} \, q_{m,i}^2 \right\} \in \Z. 
\label{E:C}
\end{equation}
This is the natural generalization of the condition $\alpha=2$ we had obtained by considering simple Kerr--Newman geometries carrying a single unit of electric charge. If we apply the Dirac quantization condition to each of these $U(1)$ fields this can equivalently be rewritten as
\begin{equation}
{1\over4} \; \sum_i  \left\{ 2\alpha_i \, q_i^2 + {q_{m,i}^2\over2\alpha_i} \right\} \in \Z. 
\label{E:C2}
\end{equation}
But now, if ordinary elementary particles (electrons, protons, neutrons) do \emph{not} carry any of these exotic extra charges, then astrophysical black holes formed from stellar collapse will never satisfy this condition. For astrophysical black holes to satisfy condition (\ref{E:C}), or equivalently (\ref{E:C2}), which is needed if equation (\ref{E:2}) is to have any chance at having universal validity, one must postulate that ordinary elementary particles (electrons, protons, neutrons) carry multiple exotic and otherwise unobserved $U(1)$ charges.  Furthermore if the fine structure constants $\alpha_i$ are generic real numbers then condition (\ref{E:C2}) implies a quite remarkable fine-tuning for astrophysical black holes, while if these fine structure constants are rational numbers one is back to dealing with the issues previously discussed. Condition (\ref{E:C2}) would also seem to imply significant constraints on both the quantity and types of elementary particle that could be ``permitted'' to undergo collapse to form an astrophysical black hole. 

This is an extremely high ``overhead'' to pay to enforce universal integer-valued validity of equation (\ref{E:2}). In conclusion, insofar as one restricts attention to simple and straightforward physical models based on the Kerr--Newman spacetimes one does not satisfy the quantization conjectures of equations (\ref{E:1})--(\ref{E:3}) in any simple integer-based manner.

%---------------------------------------------------------------------------------------------------------------------------------------------

%---------------------------------------------------------------------------------------------------------------------------------------------

\begin{thebibliography}{99}
%---------------------------------------------------------------------------------------------------------------------------------------------

\bibitem{Horowitz:1996}
  G.~T.~Horowitz and A.~Strominger,
  ``Counting states of near extremal black holes'',
  Phys.\ Rev.\ Lett.\  {\bf 77} (1996) 2368
  [hep-th/9602051].
  
  \bibitem{KeskiVakkuri:1996}
  E.~Keski-Vakkuri and P.~Kraus,
  ``Microcanonical D-branes and back reaction'',
  Nucl.\ Phys.\ B {\bf 491} (1997) 249
  [hep-th/9610045].
  %%CITATION = HEP-TH/9610045;%%
  
  \bibitem{Horowitz:1996b}
  G.~T.~Horowitz, J.~M.~Maldacena and A.~Strominger,
  ``Nonextremal black hole microstates and $U$ duality'',
  Phys.\ Lett.\ B {\bf 383} (1996) 151
  [hep-th/9603109].
  %%CITATION = HEP-TH/9603109;%%
  
  \bibitem{Halyo:1996}
  E.~Halyo, B.~Kol, A.~Rajaraman and L.~Susskind,
  ``Counting Schwarzschild and charged black holes'',
  Phys.\ Lett.\ B {\bf 401} (1997) 15
  [hep-th/9609075].
  %%CITATION = HEP-TH/9609075;%%
  
  \bibitem{Horowitz:1996c}
  G.~T.~Horowitz and J.~Polchinski,
  ``A Correspondence principle for black holes and strings'',
  Phys.\ Rev.\ D {\bf 55} (1997) 6189
  [hep-th/9612146].
  %%CITATION = HEP-TH/9612146;%%

\bibitem{Larsen:1997}
  F.~Larsen,
  ``A String model of black hole microstates'',
  Phys.\ Rev.\ D {\bf 56} (1997) 1005
  [hep-th/9702153].
  %%CITATION = HEP-TH/9702153;%%
  
\bibitem{Cvetic:1997}
  M.~Cvetic and F.~Larsen,
  ``General rotating black holes in string theory: Grey body factors and event horizons'',
  Phys.\ Rev.\ D {\bf 56} (1997) 4994
  [hep-th/9705192].
  %%CITATION = HEP-TH/9705192;%%
  
  \bibitem{Cvetic:2009}
  M.~Cvetic and F.~Larsen,
  ``Greybody Factors and Charges in Kerr/CFT'',
  JHEP {\bf 0909} (2009) 088
  [arXiv:0908.1136 [hep-th]].
  %%CITATION = ARXIV:0908.1136;%%

\bibitem{Cvetic:2010}
  M.~Cvetic, G.~W.~Gibbons, and C.~N.~Pope,
  ``Universal Area Product Formulae for Rotating and Charged Black Holes in Four and Higher Dimensions'',
  Phys.\ Rev.\ Lett.\  {\bf 106} (2011) 121301
  [arXiv:1011.0008 [hep-th]].
  %%CITATION = ARXIV:1011.0008;%%
  
  \bibitem{Galli:2011}
  P.~Galli, T.~Ortin, J.~Perz and C.~S.~Shahbazi,
  ``Non-extremal black holes of $N=2$, $d=4$ supergravity'',
  JHEP {\bf 1107} (2011) 041
  [arXiv:1105.3311 [hep-th]].
  %%CITATION = ARXIV:1105.3311;%%
  
  \bibitem{Meessen:2012}
  P.~Meessen, T.~Ortin, J.~Perz and C.~S.~Shahbazi,
  ``Black holes and black strings of $N=2$, $d=5$ supergravity in the H--FGK formalism'',
  arXiv:1204.0507 [hep-th].
  %%CITATION = ARXIV:1204.0507;%%
  
\bibitem{Castro:2012}
  A.~Castro and M.~J.~Rodriguez,
  ``Universal properties and the first law of black hole inner mechanics'',
  arXiv:1204.1284 [hep-th].
  %%CITATION = ARXIV:1204.1284;%%
  
\bibitem{Newman:1965a}
Ezra Newman and Allen Janis, ``Note on the Kerr Spinning-Particle Metric'', Journal of Mathematical Physics {\bf 6} (1965) 915--917.  doi:10.1063/1.1704350.

\bibitem{Newman:1965b}
Ezra Newman, K.~Chinnapared, A.~Exton, A.~Prakash, and R.~Torrence,  ``Metric of a Rotating, Charged Mass'', Journal of Mathematical Physics {\bf 6} (1965) 918--919.  doi:10.1063/1.1704351.

\bibitem{MTW}
  C. W. Misner, K. S. Thorne, and J. A. Wheeler, 
  \emph{Gravitation},  (W.~H.~Freeman and Company, New York, 1973).

\bibitem{Wald}
R.~M.~Wald, \emph{General relativity}, (Chicago University Press, 1984). 

 
 \bibitem{Eddington} 
 A. Eddington, \emph{Fundamental Theory}, (Cambridge University Press, 1946). 
 

\bibitem{Kerr:1963}
R.~P.~Kerr,  ``Gravitational field of a spinning mass as an example of algebraically special metrics'', Physical Review Letters {\bf 11}  (1963) 237--238.  doi:10.1103/PhysRevLett.11.237.

\bibitem{Kerr:book}
 D.~L.~Wiltshire, M.~Visser, and S.~M.~Scott (editors),
  ``The Kerr spacetime: Rotating black holes in general relativity'',
  (Cambridge University Press, 2009).
  
\bibitem{Kerr:survey}
M.~Visser,
  ``The Kerr spacetime: A Brief introduction'',
  arXiv:0706.0622 [gr-qc].
  Published in~\cite{Kerr:book}.
  %%CITATION = ARXIV:0706.0622;%%

  

\bibitem{Ansorg:2010}
  M.~Ansorg, J.~Hennig, and C.~Cederbaum,
  ``Universal properties of distorted Kerr-Newman black holes'',
  Gen.\ Rel.\ Grav.\  {\bf 43} (2011) 1205
  [arXiv:1005.3128 [gr-qc]].
  %%CITATION = ARXIV:1005.3128;%%
  
  \bibitem{Ansorg:2009}
  M.~Ansorg and J.~Hennig,
  ``The Inner Cauchy horizon of axisymmetric and stationary black holes with surrounding matter in Einstein-Maxwell theory'',
  Phys.\ Rev.\ Lett.\  {\bf 102} (2009) 221102
  [arXiv:0903.5405 [gr-qc]].
  %%CITATION = ARXIV:0903.5405;%%
  
  \bibitem{Ansorg:2008}
  M.~Ansorg and J.~Hennig,
  ``The Inner Cauchy horizon of axisymmetric and stationary black holes with surrounding matter'',
  Class.\ Quant.\ Grav.\  {\bf 25} (2008) 222001
  [arXiv:0810.3998 [gr-qc]].
  %%CITATION = ARXIV:0810.3998;%%
  
  \bibitem{Hennig:2009}
  J.~Hennig and M.~Ansorg,
  ``The Inner Cauchy horizon of axisymmetric and stationary black holes with surrounding matter in Einstein-Maxwell theory: Study in terms of soliton methods'',
  Annales Henri Poincare {\bf 10} (2009) 1075
  [arXiv:0904.2071 [gr-qc]].
  %%CITATION = ARXIV:0904.2071;%%
  
\bibitem{Planck:1899}
Max Planck, ``\"Uber irreversible Strahlungsvorg\"ange'',  Sitzungsberichte der K\"oniglich Preussischen Akademie der Wissenschaften zu Berlin {\bf 5} (1899) 440--480. Pages 478--480 contain the first appearance of the Planck base units (other than the Planck charge), and of Planck's constant, which Planck denoted by $b$. The quantities $a$ and $f$ in this paper are now typically denoted $k$ and $G$.  

  
\bibitem{Stoney:1881}
G.~Stoney, ``On The Physical Units of Nature'', Philosophical Magazine {\bf11} (1881) 381--391.
  
\bibitem{Stoney:1883}
G.~Stoney, ``On The Physical Units of Nature'', The Scientific Proceedings of the Royal Dublin Society, {\bf3} (1883) 51--60.
  

  
 \bibitem{Bekenstein:1973}
  J.~D.~Bekenstein,
  ``Black holes and entropy'',
  Phys.\ Rev.\ D {\bf 7} (1973) 2333.
  %%CITATION = PHRVA,D7,2333;%%
  
  \bibitem{Bekenstein:1974}
  J.~D.~Bekenstein,
  ``Generalized second law of thermodynamics in black hole physics'',
  Phys.\ Rev.\ D {\bf 9} (1974) 3292.
  %%CITATION = PHRVA,D9,3292;%%

  \bibitem{Dirac}
  Paul Dirac, ``Quantised singularities in the electromagnetic field'', Proc. Roy. Soc. (London) {\bf A 133} (1931) 60. 

  
%--------------------------------------------------------------------------------------------------------------------------------------------- 
%--------------------------------------------------------------------------------------------------------------------------------------------- 
%--------------------------------------------------------------------------------------------------------------------------------------------- 
%--------------------------------------------------------------------------------------------------------------------------------------------- 
\end{thebibliography}
\end{document}